\documentclass[graybox]{svmult}

\usepackage{type1cm}        
\usepackage{makeidx}         
\usepackage{graphicx}        
\usepackage{multicol}        
\usepackage[bottom]{footmisc}

\usepackage{newtxtext}       %
\usepackage{newtxmath}       

\makeindex             

\usepackage[
    backend=biber,natbib,
    style=nature,
    doi=true, isbn=false, url=false, eprint=false
]{biblatex}

\usepackage{xr-hyper}
\usepackage[breaklinks]{hyperref}
\makeatletter
\newcommand*{\addFileDependency}[1]{
  \typeout{(#1)}
  \@addtofilelist{#1}
  \IfFileExists{#1}{}{\typeout{No file #1.}}
}
\makeatother

\begin{document}

\title*{Machine Learning for Protein Engineering}
\author{Kadina E. Johnston, Clara Fannjiang, Bruce J. Wittmann,  Brian L. Hie, Kevin K. Yang, and Zachary Wu*}
\institute{Kadina E. Johnston \at California Institute of Technology
\and Clara Fannjiang \at University of California, Berkeley
\and Bruce J. Wittmann \at work done while at California Institute of Technology, now at Microsoft
\and Brian L. Hie \at Stanford University
\and Kevin K. Yang \at Microsoft Research New England
\and Zachary Wu \at \textit{Correspondence:} molzacharywu@gmail.com, now at DeepMind
\and *\textit{All authors contributed equally}
}
%
\maketitle

\begin{flushright}
\textit{This article has been accepted as an upcoming book chapter published by Springer Nature. It appears as in its original submission on February 28, 2022.}
\end{flushright}

\hfill \break
\abstract{Directed evolution of proteins has been the most effective method for protein engineering. However, a new paradigm is emerging, fusing the library generation and screening approaches of traditional directed evolution with computation through the training of machine learning models on protein sequence fitness data. This chapter highlights successful applications of machine learning to protein engineering and directed evolution, organized by the improvements that have been made with respect to each step of the directed evolution cycle. Additionally, we provide an outlook for the future based on the current direction of the field, namely in the development of calibrated models and in incorporating other modalities, such as protein structure.}

\hfill \break
Proteins are the molecular drivers for many of the processes necessary for life, performing functions such as binding small molecules, stabilizing other proteins, or catalyzing vital reactions. They are synthesized as linear sequences of individual building blocks called amino acids, of which there are twenty canonical types; these sequences fold into complex and dynamic three-dimensional structures to accomplish their functions. These sequences undergo mutation and selection based on the hosts’ abilities to reproduce.

Protein engineers often optimize proteins through a process known as directed evolution. Directed evolution adapts nature’s evolutionary process to enable the rapid improvement of proteins for human-desired functions \cite{arnold1998design, romero2009exploring}. Specifically, directed evolution iterates between modifying proteins at the sequence level and identifying protein sequences with improved level of function. This process has yielded many impactful engineered proteins, ranging from antibodies for combating disease to enzymes capable of catalyzing non-natural reactions, and its success has been recognized by the 2018 Nobel Prize in Chemistry.

Fundamentally, directed evolution is an optimization process for protein sequences. 
As such, many recent advances in machine learning have been successful in improving directed evolution’s ability to identify better proteins. 
In this chapter, we review the core concepts that have enabled successful integration of machine learning in protein engineering by interpreting the process through the directed evolution cycle.
We begin with a brief background on directed evolution, outlining the steps involved and casting them in the context of navigating protein fitness landscapes.
We then highlight recent developments in machine learning’s contribution to each of these steps as well as the experimental constraints it must operate within, before ending on an outlook for the field.
\section{Background on Directed Evolution}
\label{sec:background_theory}

Directed evolution is an adaption of natural (and artificial) selection to the modern laboratory. First, consider a simplified view of how a new protein or function may arise naturally. In nature, evolution occurs within populations and is largely driven by competition to survive, grow, and reproduce. Therefore, an attribute which confers an advantage for any of these processes is selected for, enabling an organism to out-compete the organisms around it.
The advantage could come from a mutation in a protein sequence that imparts a low level of activity for a new, beneficial function, making the organism and its offspring more successful. Thus, this protein sequence becomes more common as well. Through many iterations of this process, that protein sequence and its function could be further optimized. This process is similar to many engineering strategies, but contains notable distinctions for proteins, discussed in the following Sections \ref{subsec:DE_theory} and \ref{subsec:fitness}.

\subsection{Directed Evolution Theory}
\label{subsec:DE_theory}

The number of possible protein sequences is immense, presenting a significant challenge to any protein engineering strategy. A typical protein is several hundred amino acids long, and at each position there are twenty canonical amino acids possible. This results in \(20^{300}\) (\(10^{390}\)) possible sequences, which is around \(10^{300}\) times more than the total number of atoms in the universe. There will never be enough resources to synthesize all of these sequences let alone screen them.  

Interestingly, this does not mean that finding functional proteins or improving protein fitness is a hopeless endeavor. Maynard Smith made the critical observation that for natural selection to be possible, functional protein sequences must neighbor other functional sequences in protein space, implying that functional proteins exist clustered together within a vast sea of non-functional ones \cite{maynard1970}. Therefore, rather than throwing metaphorical darts at this astronomical space and hoping for a hit, protein engineers can begin with a small level of activity for their function of interest and leverage methods for local exploration to improve it.

\subsubsection{Protein fitness landscapes and epistasis}
\label{subsec:fitness}

Due to the iterative nature of the optimization cycle that is directed evolution, it is commonly conceptualized as a greedy, uphill walk on a \textbf{protein fitness landscape} towards a fitness peak \cite{romero2009exploring}. 
Each round of mutagenesis and screening searches through the local landscape, typically sampling only a few mutations away from the current position in sequence space. When a hit is identified, a step up the fitness peak is taken and the local search is repeated, with the entire process continuing until the fitness is satisfactory or a peak is reached. Importantly, no downward steps into valleys of the fitness landscape are typically allowed in directed evolution.

Fitness landscapes are often visualized as smooth, easy-to-navigate surfaces, but in reality they are discrete, high-dimensional spaces, with many of the dimensions being quite rugged. This ruggedness is due to a phenomenon in biology called epistasis, where mutational effects are dependent on higher order interactions rather than their individual contributions \cite{starr2016epistasis}. Epistasis arises most commonly from direct structural contacts, but interactions between residues can also be modulated by ligands, substrates, allostery, cofactors, or conformational dynamics \cite{miton2021epistasis}. As a result, it is often reasonable to assume that distant mutations are independent, but there are important cases where this assumption breaks down and epistasis must be considered.

It is also important to consider the prerequisite protein properties that must be satisfied to take fitness measurements, such as expression, stability, and substrate binding. This means that fitness landscapes are a result of some combination of these factors, and changes in any of them can modulate fitness or cause epistasis. For example, \citeauthor{romero2009exploring} outline how epistasis can arise from a protein stability threshold, where beneficial, but destabilizing mutations combine to completely ablate activity \cite{romero2009exploring}. 
The protein fitness landscape is a useful analogy for both intrinsic proteins as well as the methods in this chapter, and we come back to this concept often.

\subsection{Chapter Scope}
\label{subsec:scope}
In this chapter, we focus on methods employing machine learning for the directed evolution of proteins.
However, there are many other aspects of protein modeling that can be explored in biology \cite{ching2018opportunities} and chemistry \cite{coley2019autonomous}.
Additionally, proteins do not function in a vacuum. Other molecules, such as nucleotides that encode our genomes \cite{zou2019primer, eraslan2019deep} and small molecules \cite{vamathevan2019applications}, can have profound effects on proteins' biological functions. 

Another view of understanding proteins is through their three-dimensional structures \cite{gao2020deep}. 
Protein dynamics \cite{noe2020machine} and structure prediction \cite{alquraishi2021machine} are two rapidly growing fields, but the connection to protein function is still opaque \cite{greslehner2018molecular}.
Nevertheless, the goal of protein structure-based design \cite{ovchinnikov2021structure} is to replace much of the work required for protein engineering, instead generating optimized proteins in a single step.
However, current best efforts from this approach still benefit from subsequent rounds of directed evolution.
In the next section, we describe how machine learning methods can improve each step in directed evolution.

\subsection{Protein engineering and directed evolution}
\label{subsec:DE_cycle}

In this chapter, we view the protein engineering process through the directed evolution cycle, which adapts natural and artificial selection to the modern laboratory.
In this process, the desired properties of engineered proteins are directly queried (assayed) by man-made techniques, instead of interpreted through survival and reproduction (as in natural selection). This allows the engineer to rapidly evolve proteins for desired properties (such as thermostability or expression) and functions (such as binding or enzymatic activity).

\begin{figure}[!h]
	\centering
	\makebox[\textwidth][c]{
	    \includegraphics[height=0.8\textheight,
                          keepaspectratio=True]
                          {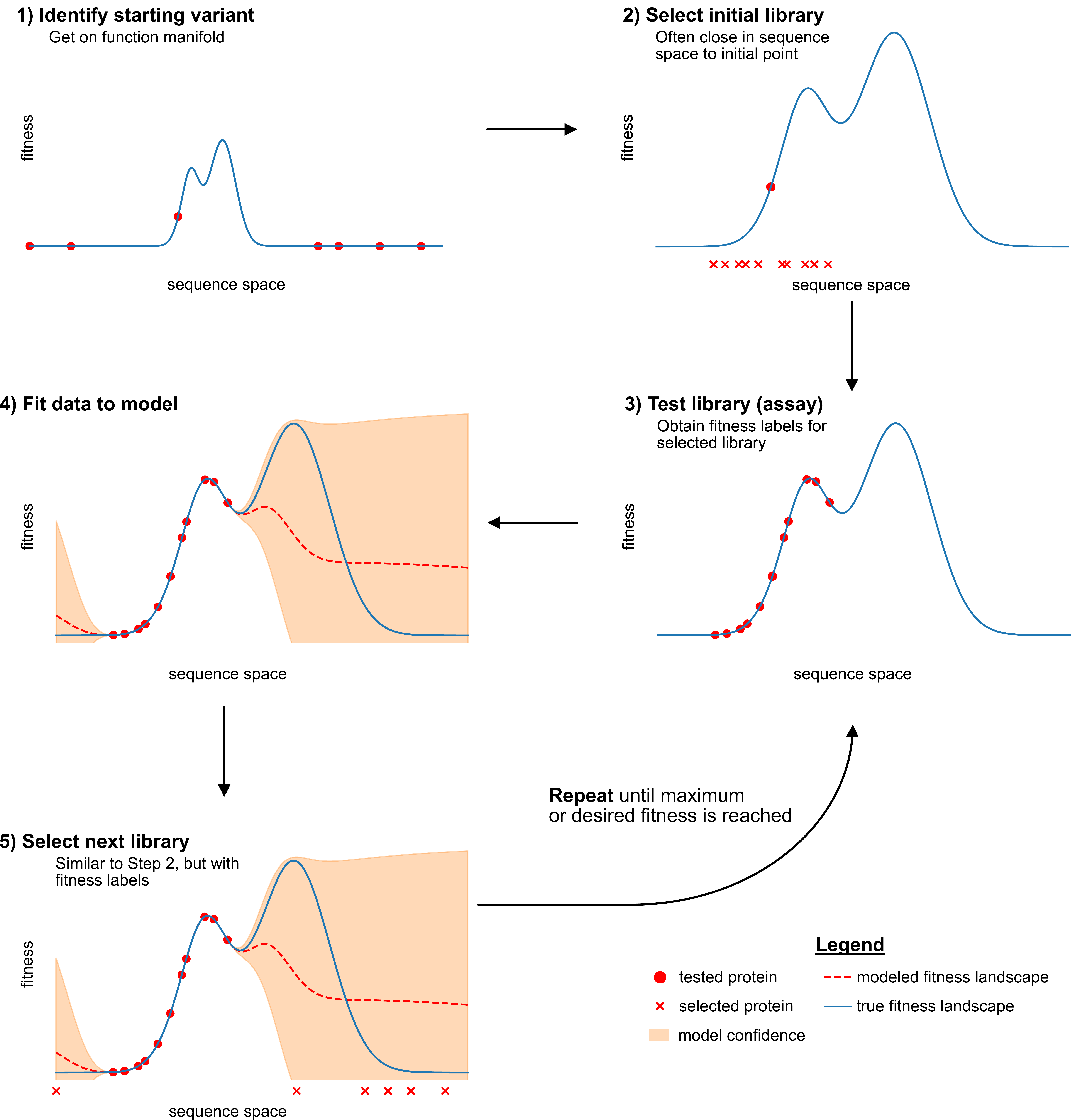}
                        }
	\caption{Protein engineering steps viewed as steps in exploring a 1D fitness landscape. In each plot, the protein fitness (scalar measurement of a function or property), is on the y-axis, and the protein sequence is on the x-axis. In (1), a protein with measurable function is identified. In (2), an initial library of protein variants is built. In  (3), the library is tested (assayed) for desired property. The next steps are additional for protein engineering methods guided by machine learning, including (4) fitting the models and (5) selecting the next library based on these models. Finally, the process is repeated until a maximum is reached, or the desired fitness satisfies a desired threshold.}
	\label{fig:de_cycle}
\end{figure}

The directed evolution cycle (Figure \ref{fig:de_cycle} can be segmented into three basic steps, with two additional steps in cases where protein engineering is driven by machine learning.
In Section~\ref{subsec:in_vitro} of this chapter we elaborate on the experimental constraints and ML considerations of these steps, and in Section~\ref{sec:ML_contributions}, we highlight examples where machine learning has enabled advances in each of the following steps:

First, a protein with a measurable amount of the property of interest must be identified. Directed evolution relies on accurate physical interrogation of this scalar value, termed fitness, and a measurable starting point is required to begin this process. Many protein engineering projects fail at this stage, as the most biochemical ingenuity is required to identify a protein that is able to accomplish often novel goals such as breaking down polyethylene terephthalate (in man-made plastics \cite{austin2018characterization}) or to selectively modify DNA (for targeted gene therapy \cite{waehler2007engineering}). 

Second, the protein sequence is randomized to generate a pool of variants, often called a \textbf{library}. In nature, this process typically occurs through random mutagenesis, where random mutations in DNA correspond to random changes in the protein sequence, or recombination of existing protein fragments. 
In the laboratory setting, current experimental techniques give protein engineers more freedom in selecting which sequences to test. One option lies in biasing the randomness of these mutations, but rapidly decreasing DNA synthesis costs also enable the generation of any specific sequence the engineer designs, such as chimeras (recombined fragments of existing proteins)~\cite{smith2013chimeragenesis, endelman2004site} or sequences generated by language models~\cite{madani2021deep,shin2021protein}. While DNA costs are currently dropping, this step can still be the limiting cost in testing more diverse protein sequences.

Third, the library is tested for the desired property. This step captures a wide range of assays that have been developed to probe biomolecules. Some examples of approaches capable of generating the largest amounts of data (high-throughput assays), are based on protein fluorescence or binding, and reach hundreds of thousands of labels per month. However, other properties such as enzymatic activity for generating small molecule substrates are measured in much lower throughput, typically on the scale of hundreds per month. This step often limits the \textit{amount} of data labels the machine learning modeler can expect to obtain.

In traditional directed evolution, the top $k$ variants (often $k=1$) are fed back into the second step for further improvement. For methods where directed evolution is further enabled by machine learning, two other steps may follow.

In these machine-learning-assisted approaches, the next step is to fit models to the relationship between proteins and their fitness labels. A wide variety of approaches are available to the machine learning practitioner here, and there are multiple sources of prior knowledge that can be leveraged for proteins. One example is the rich historical record of protein sequences, which can be obtained from sequence databases such as UniRef \cite{suzek2007uniref, suzek2015uniref}. From such a database, sets of evolutionarily-related sequences (\textbf{homologs}) can be obtained and aligned in Multiple Sequence Alignments (\textbf{MSA}s), which can be used as priors on viable sequences. However, while this history represents sequences retained in nature, it does not necessarily represent the distribution of allowed sequences for a specific protein on an engineered task. 

Finally, these models are used to select optimal proteins for experimental validation. Again, a variety of approaches have been employed in this step such as gradient-based, RL-inspired, and active learning methods (Section \ref{subsec:model_update}). The sampling strategy often depends on the modeling approach used in the previous step. Additionally, this step is similar to the second in that both may be constrained by current molecular biology techniques. These constraints are often enforced manually, but can be encoded in the design process as well. From here, directed evolution re-enters the third step, and the cycle is repeated until the optimum is reached or design criteria are satisfied.

The process of directed evolution with machine learning is depicted in Figure \ref{fig:de_cycle}, which interprets each step in the analogy of climbing protein fitness landscapes.

\subsection{Evolutionary Optimization Methods}
\label{subsec:evolutionary_optimization}
Evolution is the core optimization process in biology. Naturally, it has inspired other optimization algorithms, particularly in the gradient-free or black box optimization settings. Here, we briefly discuss the key similarities and distinctions between natural evolution, directed evolution, and evolutionary algorithms.

Natural evolution occurs on the level of organisms interacting with their environment. Mutations fixed on the individual protein level may not have a noticeable effect on fitness at the organism level, defined as its ability to survive and propagate its genes. Additionally, the objective function of natural evolution is a competition for scarce resources in constantly changing environments. Diverse solutions are often implicitly rewarded for their ability to access new ecosystems. However, for the protein engineer with the goal of optimizing individual molecules, there is not a clear mapping from the evolutionary history of similar sequences to fitness on the protein level, defined as how well a protein achieves its target function.

Evolution strategies at the protein level are the subject of this review. Two key differences distinguish directed evolution of proteins from natural evolution. In traditional directed evolution campaigns, only a single parent protein sequence is used to start each round of subsequent evolution, in contrast to nature which evolves with populations. Recently, approaches where directed evolution is further guided by machine learning captures information from all explored variants in trained, surrogate models of the protein fitness landscape \cite{yang2019machine}, which can also explicitly reward diverse solutions \cite{linder2020generative}.

Additionally, fitness is more well-defined in directed evolution than in natural evolution, where fitness is influenced by a complex set of factors and can change based on environment. For example, the fluorescence of green fluorescence protein may be directly measured, and variants may be evolved for brighter fluorescence or alternative emission wavelengths. While this is largely an advantage, it is important to remember that the process of sample preparation can also affect the measurement and therefore unintentionally modify the definition of ``fitness'' being optimized. For example, the measured binding efficiency of a protein can also depend on the purification and expression method used to obtain it. In other words, a protein must pass both purification and expression to bind effectively to a target to be considered ``fit'' in this case. This concept is captured in the directed evolution adage: "You get what you screen for" \cite{arnold1998design}.

Evolutionary algorithms are a class of optimization algorithms that are directly inspired by natural evolution, and a subset of these methods have direct analogs in directed evolution methods \cite{back1996evolutionary, beyer2001theory}. 
For example, genetic algorithms also define fitness as the value of the objective function to be optimized.
During optimization, a population of individuals (variants) is evaluated against this fitness function, and the surviving population of solutions is allowed to recombine and mutate for the next generation of individuals.
While the details of genetic selection and mutation differs from that of natural evolution, the core principles governing both processes enable successful optimization, often with diverse final solutions.
Additionally, as these methods are computational and act on digital fitness landscapes, digital evolution is more parallelizable and can be much easier to study.
However, these strategies are also subject to the same pitfalls as directed evolution methods. Interestingly, a common failure mode shared with directed evolution is in specification of the fitness landscape: computational agents often exploit bugs in the modeled physical environments to achieve the metrics \cite{lehman2020surprising}. Evolutionary algorithms can also use explicit probability distributions to generate proposal sequences, which are closely related to Monte Carlo Expectation Maximization methods \cite{brookes2020view}.

\section{ML Contributes to many steps in directed evolution}
\label{sec:ML_contributions}

The directed evolution cycle powered by machine learning methods can be segmented into five steps, as described previously. In this section, we summarize important contributions machine learning has made, and promises to make, to improve over the directed evolution baseline over each of its steps.

\subsection{Identifying starting variants}
\label{subsec:starting}

The first consideration in a directed evolution experiment is selecting the protein variant to evolve, which can be a nontrivial task.
In the most extreme setting, there may be no known proteins that perform the desired function; a more common but related setting is when the desired output of the directed evolution experiment needs to be substantially different from all known proteins due to considerations of scientific novelty or intellectual property.

In such settings, machine learning can be a useful tool to help identify a sufficiently novel starting variant, typically with weak or suboptimal fitness, that can subsequently be given to a traditional or machine-learning-guided directed evolution pipeline.
The two main approaches for doing so are (1) to screen a large and diverse collection of proteins for any that have a nonzero fitness value~\cite{shin2021protein,liu2020antibody,madani2021deep} or (2) to de-novo design the initial, functional protein~\cite{cao2021robust,huang2016coming,gligorijevic2021function}.
Machine learning has contributed to both approaches, particularly for the design of protein-protein binding.

As an example of the first approach, \citeauthor{shin2021protein} use an autoregressive language model, trained on approximately 1.2 million natural llama nanobody sequences, to generate a nanobody library that is screened to potentially identify novel binders to a target protein \cite{shin2021protein}.
The generative model enables improved sequence diversity over previous synthetic libraries, enabling the authors to identify new proteins with high binding affinity using an efficient set of approximately $10^5$ generated sequences, which is 1000-fold smaller than other libraries.

As an example of the second approach, \citeauthor{cao2021robust} de-novo design miniproteins that bind to a known, target structure \cite{cao2021robust}.
They use a multistep computational pipeline, which includes supervised machine learning models that guide physics-based approaches to protein design, to engineer de-novo binders with some affinity for a target protein.
These initial binders are then improved via traditional directed evolution to produce the final binder.

While we have mostly considered cases in which the starting variant is unknown or must be substantially different from existing proteins, this is not a constraint in many other settings where a good, functional variant exists and the goal is simply to improve fitness.
Even in such settings, however, directly evolving the best-known variant may not be the most effective strategy.
A known, high-fitness variant may be in a \textbf{local} optimum of the fitness landscape such that the evolutionary path to the unknown \textbf{global} optimum (or the highest-fitness variant) requires crossing a \textbf{fitness valley} (where some intermediate mutations along the path decrease fitness).

Therefore, it may be desirable to begin evolution at a lower-fitness variant from which multiple fitness peaks (both locally- and globally-optimal) are accessible via a greedy uphill walk strategy typical in protein engineering.
Engineering a more \textbf{evolvable} starting variant, even at some fitness cost, is currently an open question~\cite{trudeau2018protein} for which machine learning could play an important role.
Proteins with high intrinsic stability are thought to be more evolvable~\cite{bloom2006protein}, and many machine learning models have been developed that either directly (via supervision) or indirectly (as an emergent property of an unsupervised model) predict stability~\cite{cao2019deepddg,li2020predicting,meier2021language,hie2022evolocity}.
Moreover, we note that directed evolution that begins at a local optimum could still cross fitness valleys via more complex, model-guided strategies beyond greedy exploration, which we review in the following sections.

\subsection{Building initial library}
\label{subsec:library_build}

The data used to train a machine learning model determines what it learns and, by extension, in what situations it can be used to make effective predictions. 
For protein engineering, this means that the design of the library that will provide training data is critical to the eventual effectiveness of the trained model in finding improved sequences. 

Machine learning models tend to be more effective at interpolation than extrapolation and so will typically perform best when used to make predictions in the same domain as the data used to train them. 
In general, for a given \textbf{design space} of allowed proteins, this translates to collecting maximally diverse training data that best covers that space. 
For proteins, this means that training data with maximal sequence diversity will be most informative for modeling an underlying true fitness landscape: \footnote{Strictly speaking, training data that is maximizes diversity in the \textit{encoded} sequences will be the most informative} the more diverse the training sequences are, the more of the design space that is covered by the training data and the less a model must extrapolate to previously unseen regions of sequence space. 
For example, \citeauthor{romero2013navigating}, \citeauthor{bedbrook2017machine}, and \citeauthor{greenhalgh2021machine} maximize the information entropy of the initial set of sequences when engineering P450s, channelrhodopsins, and acyl-ACP reductase, respectively \cite{romero2013navigating, bedbrook2017machine, greenhalgh2021machine}. 

Randomly collecting sequences from a fixed design space (e.g., a combinatorial space defined by a given number of positions in a protein) can thus be a valuable strategy for training data collection, as this will on average result in the collection of highly diverse sequences. 
Random collection of training data is also an attractive approach from an applications point of view, as fully random libraries can be easily constructed in a multiplexed fashion using degenerate oligomers or strategies like error-prone Polymerase Chain Reaction (epPCR)~\cite{kille2013reducing} (see Section \ref{subsec:library_methods} for more information), and this strategy has been combined with machine learning to engineer halohydrin dehalogenase~\cite{fox2007improving}, fluorescent proteins~\cite{saito2018machine,somermeyer2021heterogeneity}, and an adenovirus capsid protein~\cite{bryant2021deep}.

While building a perfect map of a fitness landscape would be ideal for model-guided engineering, it is not always feasible given our limited ability to collect experimental data. 
More complex fitness landscapes considering larger sections of sequence space require more data to model and a small amount of randomly selected training data may be spread too thinly across the design space to build a comprehensive map \cite{brookes2022sparsity}. 
The goal of ML-assisted protein engineering is not to comprehensively map fitness landscapes, but to use ML to guide exploration of fitness landscapes to reach higher-fitness protein variants, as discussed further in Section \ref{Modeling the sequence-fitness relationship}.
As a result, if training data is expensive to collect, then it can be advantageous to build \textbf{focused} initial libraries that are biased toward protein variants believed \textit{a priori} to be higher in fitness. 
It is more important to be able to identify the highest-fitness variants from the set of high-fitness variants than the lowest-fitness variants from the set of low-fitness variants, and so the idea of this strategy is to model (potentially) higher-fitness regions of the protein fitness landscape at higher resolution and lower-fitness regions of the protein fitness landscape at lower resolution. 

Focused libraries can be particularly helpful when navigating hole-filled protein fitness landscapes~\cite{wittmann2021informed}. 
As more mutations are made to a protein, the probability that it retains function decreases exponentially \cite{bloom2006protein}, and so fitness landscapes consisting of combinations of mutations at multiple positions (\textbf{combinatorial} landscapes) tend to be dominated by proteins with zero or extremely low fitness. 
These variants are commonly referred to as \textbf{holes} in the fitness landscape as they only provide information on which mutations destroy protein fitness. These variants are conceptually distinct from fitness \textbf{valleys} mentioned previously, as they do not provide information about the \textit{extent} to which a mutation impacts protein fitness, which is valuable information for training the regression models typically employed for ML-assisted protein engineering. 
\citeauthor{wittmann2021informed} demonstrated that by using so-called zero-shot predictor---models or strategies that can predict protein fitness prior to collection of new experimental data---focused training sets can be constructed that minimize inclusion of holes in training data \cite{wittmann2021informed}. 
Through simulation on a complex, hole-filled, combinatorial fitness landscape, they showed that models trained with these focused training sets tended to be far more effective at identifying the highest-fitness variants than models trained with data drawn randomly from the landscape.

The prior information needed to construct focused libraries can come from many sources. For instance, prediction of protein thermal stability~\cite{wittmann2021informed}, use of meta-predictors of protein fitness~\cite{gray2018quantitative}, or strategies based on evolutionary conservation can all be used to make zero-shot predictions of protein fitness~\cite{liao2007engineering,musdal2017exploring,hopf2017mutation,riesselman2018deep,meier2021language}. 
The exact strategy that will be most effective, however, will vary depending on the fitness and protein being optimized. Take, for for instance, the zero-shot strategies that rely on sequence conservation.
Such strategies assume that evolutionary fitness aligns with whatever fitness is being predicted; that is, they assume that mutant proteins more closely resembling known protein sequences (found in databases of protein sequence such as UniProt~\cite{uniprot2021uniprot}) are more likely to be functional than others. 
Should this assumption not hold (for instance, the fitness of a protein being engineered for a new-to-nature activity may not correlate well with evolutionary fitness), or if there are simply not enough homologous protein sequences available to build an effective sequence-based zero-shot prediction model, then the zero-shot predictions are likely to be inaccurate. 
Inaccurate zero-shot predictions are unhelpful for focused library design: indeed, they may even be detrimental to effective learning by focusing training data collection on regions of the fitness landscape dominated by holes.

Ultimately, the decision between random library design and focused library design will depend on a number of factors. 
If the fitness landscape to be explored is expected to be minimally complex with few holes and large amounts of training data can be easily collected for it, then random library design is a reasonable approach, as the library itself will be simple to construct, and training data gathered from it will be sufficient to build a comprehensive map of the fitness landscape. 
If the fitness landscape to be explored is complex, full of holes, and it is challenging to gather training data for it, then focused libraries may be a viable option, particularly if high-confidence zero-shot predictions can be made for the fitness landscape. 
Such libraries may be more challenging to construct in the laboratory, but they will likely result in more efficient machine learning-guided engineering.

\subsection{Experimental considerations for applying machine learning to protein engineering \textit{in vitro}}
\label{subsec:in_vitro}

The process of directed evolution is relatively simple in theory and can also be so in practice. 
However, the process must operate within current experimental constraints imposed by a variety of factors, including available technology, cost, and time. 
Current laboratory methods for directed evolution have been reviewed extensively \cite{packer2015methods, currin2015synthetic}, so we focus here on facets that can impact the usefulness of machine learning methods.
As alluded to in the previous section, there is often a trade-off between lab-work complexity or cost and efficiency of ML-guided engineering, and we discuss these trade-offs below.

Notably, both the input protein sequences and the output experimental labels both depend on physical constraints. In the first section, we discuss different approaches to generating protein libraries. In the second section, we discuss varying experimental methods (assays) for obtaining protein labels. Finally, in the third section we discuss methods for pairing protein sequence and function information.

\subsubsection{Methods for library generation}
\label{subsec:library_methods}

Random mutagenesis is one of the most straightforward methods for creating initial sequence diversity. Errors are introduced throughout an initial DNA sequence by either randomly damaging DNA or by introducing errors during replication such as via epPCR, a process that introduces mutations randomly by increasing the error rate of the copying enzyme, the polymerase. However, it is important to note that errors introduced via ``random'' mutagenesis are not perfectly random in two major ways. First, mutations from one nucleotide to another do not occur at identical frequencies, so the original base can dictate what mutations are most likely at a given position \cite{vanhercke2005reducing}. Second, the genetic code is redundant, with the 20 canonical amino acids encoded by 61, three-nucleotide codons. Although it is possible for multiple nucleotides within a single codon to mutate simultaneously, this is rare, and generally only one nucleotide mutation occurs per codon during random mutagenesis, limiting the mutations available at the amino acid level. Importantly for ML methods hoping to build upon random mutagenesis strategies, there are very few parameters that can be tuned in this process---error rate can be modulated to shift the Poisson distribution of number of mutations, different regions of the sequence of interest can be targeted, and bias towards particular nucleotide swaps can be imposed. In comparison to other mutagenesis methods, the protein engineer has much less control over the generated library.

Targeted mutagenesis, of which the focused mutagenesis discussed in the previous section is a subset, is an alternative to random mutagenesis that affords more control over the final library. Unlike random mutagenesis, targeted mutagenesis typically assumes that either (1) specific sites in the protein sequence are important to mutate or (2) it is important to be able to access all amino acids at a given position. Site selection usually requires structural knowledge or other biochemical insights into the protein system. To perform targeted mutagenesis, degenerate oligos---mixtures of individual primers that together encode a distribution of mutations---are used to induce randomness at a specific codon/amino acid position or positions. Unlike random mutagenesis, multiple mutations can now be made within a single codon, allowing access to any amino acid if desired. The pool of degenerate oligos can also be modified to achieve a relatively even distribution across all twenty canonical amino acids \cite{kille2013reducing} or to achieve a different distribution of interest \cite{weinstein2021optimal}. Targeted mutagenesis can also be performed simultaneously at multiple sites, enabling the exploration of more complex, epistatic landscapes. By tuning the distribution of mutations encoded in such simultaneous site-saturation libraries, focused libraries such as those described in the previous section can be created. Importantly, as the desired amino acid distribution becomes more complex and more sites are mutated simultaneously, both difficulty of laboratory implementation and the cost of oligos can become untenable. There are a number of computational strategies available that can assist with such complex library designs; however, all are fundamentally limited by the set of amino acid distributions made possible by the genetic code \cite{shimko2020decode, jacobs2015swiftlib, weinstein2021optimal}. Therefore, ML methods relying on targeted mutagenesis for either training set design or evaluation of predicted designs must keep in mind these constraints. 

A final strategy for library generation is recombination, which pieces together, or ``recombines'', initial diversity into different arrangements to create new diversity. Recombination is a very broad category of diversity generation, and due to the array of recombination strategies available, we do not explain them here, but they are well reviewed by \citeauthor{packer2015methods}~\cite{packer2015methods}. The choice of recombination strategy typically relies on the type of initial diversity on hand. Such diversity could be comprised of a set of functional, \textbf{homologous} proteins, the top variants from a random mutagenesis library, or the top variants from a targeted mutagenesis library.
One recombinatorial approach of note is the use of SCHEMA libraries \cite{voigt2002protein}, where fragments of multiple parent proteins are swapped, and which has been successfully engineered with machine learning methods \cite{romero2013navigating, bedbrook2019machine}.
Importantly, some recombination strategies are quite experimentally straightforward and a single round of recombination on top variants can yield much higher improvements in fitness than a single round of random or targeted mutagenesis.
Therefore, when comparing ML methods for protein engineering to lab-only methods, using a wet-lab recombination strategy as a baseline is highly recommended when possible.

There is promise to disrupt these traditional paradigms of library generation with the advent of cheap gene synthesis technologies \cite{kosuri2014DNAsynth}. Rather than starting with an initial sequence or pool of sequences and building diversity with mutagenesis or recombination, a set of desired sequences can be synthesized directly for under \$100 per gene block. Ordering pools of sequences with targeted or random mutations is also possible, and synthesis technologies can impart more control over the final distribution. As these DNA synthesis technologies improve, ML methods for protein engineering can shake the constraints imposed by existing library construction technologies, allowing researchers to design libraries with specific distributions of mutations.

\subsubsection{Assaying protein fitness}
\label{subsec:assays}

With proteins performing such a wide variety of different functions, the protein engineering community has had to devise countless different assays to measure them all. As such, assays for protein function vary widely in both accuracy and throughput, with ranges from tens to millions of protein variants. This amount is typically dependent on the project definition of fitness. For instance, if the measurement of fitness can be directly coupled to a sequencing assay (as in deep mutational scanning), then large datasets ($10^{5}$ to $10^{6}$) can be rapidly created. Many assays for fitness, however, are limited to comparatively low-throughput chromatographic methods (e.g., HPLC, LCMS, GCMS, etc.), which rely on physical separation of a mixture through a column,  producing smaller datasets ($10^{1}$ to $10^{4}$). The amount of data available for training a model will dictate how much of sequence space can be explored and how accurate the predictions of fitness for new sequences will be. At the same time, however, the goal of ML in protein engineering is to reduce the burden of experimental screening and expedite the process, so a balance must be found between the amount of data collected and the accuracy of predictions of models trained on that data.

Additionally, one should consider where an ML method for protein engineering may have the biggest impact on efficiency, cost, and time. Protein functions that can be assayed in high-throughput provide more data for training downstream ML models; however, applying an ML-based engineering method to such a function may not impart as much benefit as it would being applied to engineering a function with a low-throughput assay. Therefore, many recent efforts for building ML methods for protein engineering have focused on the \textbf{low-N} regime, where few samples are used for training \cite{wittmann2021informed, hsu2022learning, biswas2021low, wu2019combination, bedbrook2019machine, qiu2021cluster}.

Final considerations, for both the protein engineer choosing an assay and the ML scientist choosing a dataset, are the assay noise and bias. As interest has grown in applying ML to protein engineering, method developers have sought out sequence-fitness datasets with which to benchmark their approaches. This has typically resulted in the pursuit of large datasets built from high-throughput screening methods, such as deep mutational scanning (DMS) \cite{fowler2014deep} and fluorescence-activated cell sorting (FACS) \cite{bonner1972FACS}, but inherent biases in such protein fitness datasets are left under-discussed. DMS fitness measurements can be heavily impacted by the input library and can result in highly non-uniform error across a dataset \cite{rubin2017statistical}. FACS-based assays can result in bias due to binning that occurs when gates are chosen for sorting cells \cite{trippe2022randomized}. Note that all fitness assays have their own noise and bias which should be thoroughly considered when developing an ML method on a particular dataset.

\subsubsection{Pairing sequences to fitness data through sequencing}
\label{subsec:seq-fit_pairing}

Uncommon in the broader ML discipline, datasets that result from typical protein engineering campaigns are label-rich and feature-poor, with many assayed fitnesses and relatively few variant sequences. Because only function is being optimized in directed evolution, this process does not technically require sequencing. Indeed, only the top few variants from each round are sequenced for validation in practice, and sequencing all of the variants is considered an unnecessary, unjustifiable expense. Thus, the remainder of the unimproved variants are discarded without sequencing, resulting in fitness labels that are rendered useless. Notably, this is not true for deep mutational scanning libraries, where fitness is directly coupled to sequencing, which is why many ML approaches are currently developed on these types of datasets \cite{wu2019combination, wittmann2021informed, dallago2021flip, madani2020progen, aghazadeh2021epistatic}.

Variant sequencing, especially for low-throughput assays, has traditionally been done via Sanger sequencing, but this method scales linearly with the number of variants, costing a few dollars per sequence. Depending on the assay used for evaluating protein fitness, sequencing could easily become the most expensive part of a protein engineering campaign. 

Fortunately, next generation sequencing (NGS) technologies \cite{Slatko2018NGS} have begun disrupting this paradigm. The deep mutational scanning strategies mentioned in the previous section rely on sequencing to measure fitness, meaning sequencing and assaying fitness happen simultaneously \cite{fowler2014deep}. New sequencing methods that incorporate NGS show promise to continue shifting the sequencing paradigm by spreading reads over many, multiplexed sequences \cite{Currin2019multiplexed, wittmann2021evSeq, Appel2021uPIC-M}. Currently, such methods still have limitations, ranging from the requirement to fill an entire flow cell, which can cost upwards of \$1000 per run \cite{Currin2019multiplexed, Appel2021uPIC-M}, to only sequencing short amplicons from within an entire gene \cite{wittmann2021evSeq}, which are restricted to only a few hundred amino acids, complicating its application to mutations over longer sequences. 
Nonetheless, both sequencing technologies and sequencing methods for the protein engineering community have been growing rapidly in the past few years, showing promise to address these shortcomings in the near future.

\subsection{Modeling protein fitness}
\label{subsec:modeling_fitness}
Fitness datasets obtained as described above and can be used to fit models of the sequence-fitness relationship. 
Here, we first describe the methods for representing a protein sequence before diving into the model classes used to predict fitness from sequence.

\label{subsec:seq_model}
\subsubsection{Representing protein sequences}

Proteins are variable-length sequences of twenty canonical amino acids.
Due to both the variability in sequence length and the categorical nature of amino acids, an important and long-standing problem has been how to best represent a protein sequence to facilitate machine learning.
Roughly speaking, there are two general approaches for doing this: (1) hand-crafting features based on, for example, statistical summaries or biophysical properties of a sequence or its constituent amino acids, and (2) using evolutionarily related sequences to learn likelihoods or embeddings as representations.

\textit{Statistical summaries of a sequence: higher-order terms and k-mers.} One of the simplest ways to represent a protein sequence is to concatenate one-hot encodings of the amino acid at each position in the sequence.
Despite its simplicity, even a linear model with this representation can be a surprisingly effective baseline in some settings---for example, when we focus on predictive performance on sequences near a wild type \cite{dallago2021flip, hsu2022learning}.
Amino acid similarities may also be directly encoded through matrices based on substitution probabilities of one amino acid for another \cite{henikoff1992amino}.
Beyond encoding just the amino acid identities at each sequence position, one can also encode the identities of the amino acids at pairs of structurally contacting positions \cite{romero2013navigating, bedbrook2017machine}.

Generalizing this idea further, one-hot encodings of groups of amino acids, which we call \emph{higher-order terms}---for example, the identities of the amino acids at three positions for a third-order term---have been shown to be both theoretically principled and empirically useful representations for learning protein fitness functions \cite{stadler1996landscapes,weinreich2013should,poelwijk2016context,poelwijk2019learning,zhu2021machine,aghazadeh2021epistatic,brookes2022sparsity,fannjiang2022conformal}.
In particular, it has been shown that any fitness function defined over sequences of a fixed length can be expressed exactly as a linear function of all possible higher-order terms of all orders between sequence positions \cite{stadler1996landscapes}.
The number of such terms grows exponentially with sequence length.
However, there is increasing evidence that many realistic protein fitness functions can be well-approximated by a linear model of just a small number of these terms \cite{ballal2020sparse,aghazadeh2020crisprland,brookes2022sparsity}; an effective baseline representation may therefore be to one-hot encode all terms up to, for example, second- or third-order \cite{poelwijk2019learning,hsu2022learning,fannjiang2022conformal}.

Beyond higher-order terms, the frequencies of different \emph{k-mers}---contiguous subsequences of $k$ amino acids---are another kind statistical summary that has been used to represent sequences and learn fitness functions \cite{ofer2015profet, mellor2016semisupervised}.

\textit{Physicochemical features.} Beyond one-hot encodings of a sequence's residues, or the higher-order terms between them, features based on physicochemical properties of a sequence have been used to learn fitness functions.
For example, each amino acid can be featurized as a vector of various physicochemical properties \cite{sandberg1998new, kawashima2008aaindex} or by a low-dimensional representation thereof \cite{georgiev2009interpretable, barley2018improved, tian2007tscale}.
These properties can also be combined with structural information \cite{qiu2007structural,buske2009silico, pires2013mcsm}, or with the aforementioned statistical summary representations \cite{ofer2015profet}.

However, it not generally clear which physicochemical properties will be relevant for any particular fitness function.
Consequently, in recent years considerable attention has turned to \textit{learning} relevant features by fitting low-dimensional representations of evolutionarily related sequences.

\textit{Representations learned from evolutionarily related sequences.} In directed evolution, one generally starts by identifying a \textbf{wild-type} protein (or multiple such wild types) that exists naturally in the genome of some organism and exhibits the function of interest to an appreciable degree.
In many cases, that function of interest is needed by more than one organism, and natural selection has yielded multiple homologs residing in a variety of organisms.
Though the precise fitness value differs across homologs, due to evolutionary constraints homologs are generally functional as opposed to non-functional.
Based on the premise that homologs are mostly functional, a great body of work in recent years has focused on learning the distribution of homologous sequences, as a proxy for the distribution of functional sequences.
The density, or approximations thereof, of the resulting learned sequence distribution has been shown to correlate with fitness in many cases, whether modeling the homologs with a hidden Markov model \cite{shihab2013predicting,hsu2022learning,xie2022enhancing}, Potts model \cite{mann2014fitness, hopf2017mutation}, variational autoencoder \cite{riesselman2018deep, ding2019deciphering, frazer2021disease}, or autoregressive model \cite{shin2021protein}, leading to its use as an informative correlate of fitness that does not require any experimental measurements.

Currently, variational autoencoders such as \cite{riesselman2018deep} appear to provide the highest correlations with fitness for a variety of fitness data sets, presumably due to their ability to learn evolutionary constraints on higher-order interactions between sequence positions, though autoregressive models \cite{shin2021protein} and transformers \cite{meier2021language} also achieve comparable performance.
Due to this correlation, approximations of the density such as the evidence lower bound (ELBO) have shown to be an effective feature when used with supervised learning approaches \cite{hsu2022learning}; the latent space learned by variational autoencoders has also been used to represent proteins for learning fitness functions \cite{ding2019deciphering}.
Finally, generative models fit to homologs have also been used to sample novel functional proteins \cite{russ2020evolution, madani2021deep, hawkins2020generating}.
 
Beyond homologs, which, by virtue of natural selection, contain sequence patterns necessary for the function of interest, the set of all proteins sequenced so far may similarly reveal patterns necessary for basic properties of all functional proteins, such as the ability to fold into a stable structure.
Based on this hypothesis, a growing body of work has trained various deep learning models on massive databases of known proteins, such as UniProt \cite{uniprot2021uniprot}, UniRef \cite{suzek2015uniref}, and UniParc \cite{uniprot2021uniprot}, which contain up to hundreds of millions of sequence as of this writing.
The model architectures are almost all borrowed from natural language processing (NLP), ranging from relatively simple skip-gram neural networks \cite{asgari2015continuous, ng2017dna2vec} and its extensions to embedding text documents \cite{kimothi2016distributed, yang2018learned} to recurrent \cite{alley2019unified, bepler2019learning, biswas2021low} and convolutional neural networks \cite{schwartz2018deep} to, most recently, transformers \cite{rao2019evaluating, madani2020progen, rives2021biological, rao2021msa, meier2021language, elnaggar2021prottrans}.
These models are generally trained using self-supervision tasks, where the goal is, for example, to predict the amino acid at a certain position given the amino acids at other positions (though other approaches such as multi-task learning using function labels \cite{schwartz2018deep} and incorporating protein structural information \cite{bepler2019learning} have also been used).
The outputs of intermediate layers in these models, which we will call deep embeddings, can then be used to represent protein sequences for learning fitness functions.

Estimates of the density from transformers have also been shown to correlate with a variety of fitness functions \cite{meier2021language, rao2021msa}, presumably both because the model learns sequence patterns required for basic properties of all functional proteins, such as stability, and because the training data contains all homologs corresponding to all known functions.
Whether using them for deep embeddings or density estimates, if homologs of a functional wild type are also available for a function of interest, these models can also be further trained or ``fine-tuned'' on them \cite{biswas2021low, madani2021deep, rao2021msa} to focus on a particularly relevant region of sequence space.

Despite promising results on capturing phylogenetic relationships and various clustering tasks, these deep embeddings have shown limited success in learning fitness functions.
For example, when used with ridge regression or Gaussian process regression, they do not appear to systematically outperform simpler baselines \cite{yang2018learned,hsu2022learning,dallago2021flip} such as concatenating the aforementioned density of a homolog sequence distribution to a one-hot encoding \cite{hsu2022learning}, particularly when there is limited training data.

\subsubsection{Modeling the sequence-fitness relationship}
\label{Modeling the sequence-fitness relationship}

Given a representation of protein sequences, a variety of model classes have been used to learn the sequence-fitness relationship.
In general, however, note that models that excel under more traditional evaluation metrics---for example, prediction error on held-out sequences from the same distribution as the training data---may or may not be appropriate for use with directed evolution, since in the latter we focus on sequences the model predicts to have high fitness.
That is, a model that facilitates directed evolution needs to be reliable even in regions of sequence space far from the training data, particularly in regions on which it predicts high fitness.

With that in mind, a few domain-specific models have been designed with the particular goal of jointly learning (1) a density model over homologs and (2) the fitness function from experimentally labeled sequences \cite{barrat2016improving, shamsi2020tlmutation}, under the premise that the density of homologs is positively correlated with the fitness and that therefore both models should stand to gain from sharing information via joint training.

In the more common setting of using only experimentally labeled sequences, two models have been shown to be very strong baselines across a variety of fitness data sets, particularly in the regime of limited training data (i.e., at most a few hundred labeled sequences): linear regression, for example on one-hot encodings \cite{fox2007improving, li2007diverse, dallago2021flip}, optionally concatenated with the density of a homolog sequence distribution \cite{hsu2022learning}, and Gaussian process regression with either domain-specific kernels (e.g., structurally informed \cite{romero2013navigating, jokinen2018mgpfusion, bedbrook2017machine} or mismatch string \cite{leslie2004mismatch, yang2018learned} kernels, which count shared subsequences between proteins) or more general-purpose kernels \cite{bedbrook2019machine, pires2013mcsm,mellor2016semisupervised, saito2018machine, yang2018learned, hie2021learning}.

Other kernel methods have been used as well, such as support vector machines for modeling enzyme enantioselectivity \cite{zaugg2017learning}, membrane protein expression \cite{saladi2018statistical}, and protein thermostability \cite{tian2010predicting, li2012prots, jia2015structure, capriotti2005mutant2, capriotti2005predicting, cheng2006prediction, buske2009silico, liu2012grading}.
The latter task, which has been of widespread interest due to its necessity in industrial applications, has also been approached using regression trees and their extensions \cite{tian2010predicting, li2012prots, jia2015structure}.

When labeled sequences are abundant (that is, at least tens to hundreds of thousands), deep learning models can also be effective for learning the fitness function. Models range from simple feedforward networks, which have been used to predict fluorescence \cite{sarkisyan2016local, brookes2019conditioning}, histidine synthesis efficacy \cite{pokusaeva2019experimental}, and viral capsid packaging ability \cite{zhu2021machine}, to convolutional neural networks \cite{shanehsazzadeh2020transfer, lu2020self, dallago2021flip, bryant2021deep}, recurrent neural networks \cite{rao2019tape, alley2019unified, bryant2021deep}, and transformers \cite{rao2019tape, dallago2021flip}, which have been assessed on a variety of protein tasks including fluorescence, thermostability, and binding affinity prediction.

\textit{Limitations of current model comparisons.} Our understanding of what representations and models are most effective is limited by the constraints of the molecular biology techniques available to generate fitness data sets.
For example, the vast majority of currently available fitness data sets only assess the fitnesses of mutants one or two mutations away from a wild type, excluding insertions and deletions \cite{hopf2017mutation, gray2018quantitative, riesselman2018deep, shin2021protein, hsu2022learning}.
This choice is usually justified by the assumption that most mutations are deleterious \cite{bloom2005thermodynamic}; therefore, to generate data with informative, non-zero labels, it is beneficial to stay near sequences already known to be functional. 
Pragmatically, the resource and throughput constraints on mutagenesis and sequencing protocols described in Section \ref{subsec:assays} further discourage protein engineers from making more simultaneous mutations.
If one intends on generating training data near a functional wild type, comparing representations and models using these data sets may be sufficiently informative.
However, beyond this limited setting, fitness data sets with measurements for proteins more broadly dispersed in sequence space are needed to comprehensively understand what machine learning strategies are most effective for learning fitness functions.
\citet{dallago2021flip} have collected several such datasets in recent work; there are also a few data sets with fitness measurements for all possible protein sequences varying at a small number (typically less than twenty) of positions \cite{poelwijk2019learning, wu2016adaptation, pokusaeva2019experimental}, which comprehensively expose the effects of all possible combinations of (albeit a limited number of) mutations.

Current evidence suggests that, when focusing on relatively small regions of sequence space---for example, clustered around a wild type---simple models such as ridge regression on one-hot encodings can be much more effective than more complex models \cite{dallago2021flip, hsu2022learning}, but when considering predictive performance on a broader region, given sufficient training data there may be benefits from using transformers pretrained on massive protein databases \cite{dallago2021flip}.
Further work is needed to systematically investigate the machine learning strategies appropriate for the different types of training and test data that arise in protein engineering.

\subsection{Protein selection and optimization}
\label{subsec:optimization}

After obtaining a trained sequence-fitness model, the goal is to use it as part of an \textbf{acquisition function} (in the Bayesian optimization sense) to decide which sequences from the theoretical library to characterize next. 

\subsubsection{Defining the design space}
\label{subsec:selection_designspace}
Although evaluating the sequence-fitness model is faster and cheaper than actually obtaining laboratory measurements, it is frequently impossible to evaluate it on all possible protein sequences.
One approach is to explicitly define a more limited \textbf{design space} of proteins.
Example design spaces include traditional library designs for protein engineering such as all single- or double-mutants, all the possible mutations at a small number of sites~\cite{wu2019combination,wittmann2021informed,wu2016adaptation}, a recombination library defined by mixing and matching parts of homologous parent sequences~\cite{romero2013navigating,voigt2002protein,smith2013chimeragenesis,endelman2004site}, or libraries that are easy to make using current DNA synthesis techniques~\cite{weinstein2021optimal}.
If a differentiable predictive oracle is trained on sequence-fitness data, another approach is to optimize the sequence using gradient ascent \cite{linder2020fast}.

Another approach is to use a \textit{generative model} to implicitly define the design space by learning a probability distribution over sequences.
A generative model performs one or both of two fundamental tasks: (1) assigning every possible sequence a likelihood of being in the desired distribution, and (2) generating examples of sequences from a desired distribution.
The simplest models assume this distribution can be modeled by considering sites independently or by relationships between pairs of sites~\cite{hopf2017mutation,russ2020evolution}.
More recently, researchers have used deep generative models to learn more complex sequence distributions and propose sequences for evaluation~\cite{wu2021protein}; relevant neural architectures include variational autoencoders (VAEs)~\cite{kingma2013auto,riesselman2018deep}, generative adversarial networks (GANs)~\cite{goodfellow2014generative,gupta2019feedback, repecka2019expanding}, and autoregressive language models~\cite{hochreiter1997long,bepler2019learning,shin2021protein,madani2020progen,madani2021deep}.

\subsubsection{Defining the acquisition function}
\label{subsec:acquisition_fn}
An acquisition function uses sequence information and the surrogate model to prioritize sequences from the theoretical library for experimental measurements.
A simple example of an acquisition function is greedy selection of the top prediction (or the top few predictions) according to the sequence-function model.
Greedy acquisition is common in practice and can work well~\cite{fox2007improving,wu2019combination,bryant2021deep,biswas2021low,wittmann2021informed,singer2021large}, but can become trapped in more complex fitness landscapes. 
Many standard acquisition functions are designed to select only a single example in each round.
Often, however, it is faster to obtain many experimental measurements in parallel.
While simply acquiring several of the top-ranked sequences is possible, this approach may result in the acquisition of many similar sequences.
Special methods for batched acquisition are therefore designed to encourage acquiring more diverse sequences~\cite{gonzalez2016batch,azimi2010batch,romero2013navigating,desautels2014parallelizing,yang2020batched,sinai2020adalead}, but this remains an open area of methodological development.

Given a theoretical library and an acquisition function, the simplest method for choosing sequences is to compute the surrogate score on as many members of the library as is computationally feasible, and then to use those scores to select sequences. 
However, when the library is defined by a generative model, the sequence-fitness model can be used to shift the generated distribution toward more optimal sequences.
For example, in adaptive sampling, sequences are sampled from a generative model, the outputs of a surrogate model are used to re-estimate the parameters of the generative model, and the process iterates until convergence~\cite{brookes2018design}.
\citeauthor{brookes2018design} use adaptive sampling with a VAE to optimize DNA sequences for protein expression abundance \cite{brookes2018design}.
\citeauthor{gupta2018feedback} use adaptive sampling based with a GAN to design antimicrobial peptides \cite{gupta2018feedback}.
To perform \textit{de novo} protein design \citeauthor{anishchenko2021novo} use adaptive sampling with a mutation-based generative model to identify sequences with valid folds.

By default, adaptive sampling assumes a trustworthy sequence-fitness model; however, in practice, these models are imperfect and can be prone to poor predictions in many regions of the protein space.
To address this problem, adaptive sampling can avoid these degeneracies in the surrogate model by constraining sampling to be close to the training distribution for the surrogate model~\cite{brookes2019conditioning,anishchenko2021novo}.
In each iteration, it is also possible to retrain the surrogate model to avoid pathologies~\cite{fannjiang2020autofocused}.
Sequence generation can also be improved by sampling from an ensemble of generative models~\cite{angermueller2020rmodel, angermueller2020population}.

A closely-related approach uses genetic algorithms to heuristically balance both mutation and recombination to produce new sequences~\cite{hansen2006cma} by adaptively querying a sequence-fitness model to preserve sequence designs~\cite{sinai2020adalead}.
Another approach inverts the sequence-fitness model by finding the elements from the generative model's distribution that are most likely to have a desirable value according to the sequence-fitness model.
An inverse of the sequence-fitness model can be trained via an iterative procedure similar to adaptive sampling~\cite{kumar2019model}, or the inverse of a differentiable surrogate model can be computed using gradient-based methods~\cite{liu2020antibody,linder2020generative,linder2020fast}.

While it is computationally convenient to exactly specify the sequences to measure next, specifying a new library and sampling from that library may allow many more sequences to be measured for the same cost as directly synthesizing a small number of sequences. This library may be optimized for expected improvement~\cite{yang2020batched}, to balance diversity and expected function~\cite{zhu2021machine}, or to match a desired distribution over sequences~\cite{weinstein2021optimal}. 

\subsubsection{Updating the sequence-function model}
\label{subsec:model_update}
So far, we have discussed methods for using an acquisition function based on a sequence-fitness model to choose sequences from a theoretical library that are optimized for a desired function. 
However, in some cases it may be possible to sequentially update the sequence-fitness model with new measurements. 

Greedy acquisition across experimental rounds (uphill climbing the fitness landscape) is the simplest implementation of sequential optimization and is used widely in practice~\cite{fox2007improving}.
Going beyond greedy acquisition means tolerating more risk for a potentially higher reward, which is often described as a tradeoff between \textbf{exploitation} (equivalent to greedy acquisition) and \textbf{exploration}~\cite{robbins1952some,auer2002using,snoek2012practical,sutton2018reinforcement}, in which an algorithm acquires proteins where the model is uncertain about its predictions (or that are are dissimilar from sequences in the training set) to improve future predictions and to explore new regions of sequence space.

Bayesian optimization is a popular framework for sequential optimization~\cite{snoek2012practical} that leverages Bayesian uncertainty in the sequence-fitness predictions to guide the exploration-exploitation tradeoff. 
Bayesian optimization relies on an acquisition function that systematically weighs the sequence-fitness model's prediction with its associated uncertainty.
The upper confidence bound (UCB) acquisition function adds the prediction value with a weighted uncertainty term that lets the user control the influence of uncertainty on the prediction, with a larger weight encouraging more exploration~\cite{auer2002using,srinivas2009gaussian}.
UCB has good theoretical properties~\cite{snoek2012practical} and is used widely in practice~\cite{romero2013navigating,bedbrook2017machine,greenhalgh2021machine,hie2020leveraging}.
Other notable acquisition functions select an example that is predicted to, in expectation, have the largest improvement compared to the best example in the training set or compared to a randomly drawn example from the training set~\cite{wilson2018maximizing}.
Uncertainty can also help improve exploration in parallel acquisition~\cite{azimi2010batch,romero2013navigating,desautels2014parallelizing,gonzalez2016batch,yang2020batched}.

Gaussian process models~\cite{rasmussen2006gaussian} are a popular kernel method for Bayesian optimization because of their theoretical elegance, flexibility, and good performance in practice.
However, Bayesian optimization can also leverage more bespoke Bayesian models and algorithms for exact or approximate inference.
The wide interest in neural network models over the last decade has also led to increased interest in uncertainty prediction through Bayesian neural networks, in which the parameters of the network are themselves random variables with associated prior distributions, though efficient and accurate inference in these models can be challenging~\cite{neal2012bayesian}. 
One particularly interesting approach is to combine a fitness-space prior (e.g. stability predictions) with the standard neural network weight-space prior~\cite{nisonoff2022augmenting}, and more examples are described in Section \ref{Modeling the sequence-fitness relationship}.

Probabilistic surrogate models can also be implemented by model ensembles~\cite{laks2017ensemble}, which train multiple sequence-to-fitness models on the same data and rely on variance in model predictions, due to different model architectures or randomness in the training procedure, to estimate uncertainty~\cite{liu2020antibody}.
Ensembles are not Bayesian by default, so incorporating prior information into these models can be challenging~\cite{amini2019deep,izmailov2021what}.

Finally, conformal prediction (CP) is an approach for rigorous uncertainty quantification that can be used with any predictive model in a black-box manner, including the aforementioned Gaussian process models or any deep learning model \cite{gammerman1998learning, vovk2005algorithmic, angelopoulos2021gentle}.
The framework leverages the assumption of exchangeable (e.g., independently and identically distributed) training and test data to construct confidence sets around predictions that provably satisfy \emph{coverage}, a frequentist guarantee on how often the confidence sets contain the true label of test inputs.
Beyond exchangeable data, CP has been generalized to handle various forms of distribution shift, including label shift \cite{aleks2021distribution}, covariate shift \cite{tibshirani2019conformal, park2021pac}, distribution shifts in an online setting \cite{gibbs2021adaptive}, and test distributions that are close to the training distribution \cite{cauchois2020robust}.
Notably, in recent work \citeauthor{fannjiang2022conformal} generalized CP for the setting of protein design, in which any of the aforementioned optimization procedures induces a distribution shift wherein the distribution of test (i.e., designed) sequences is dependent on the training data \cite{fannjiang2022conformal}.
This method enables practitioners to assess the predictive uncertainty of any optimization procedure, using any machine learning model, and as such can be a principled approach for selecting an optimization procedure (or any component thereof, including any hyperparameters or the machine learning model class).


\section{Outlook and Concluding Remarks}
\label{sec:future}
In this chapter, we have discussed current efforts to improve protein engineering with machine learning. The field has grown rapidly over the past five years, as the development of deep learning methods and methods to obtain and organize biomolecular data have reached inflection points. We anticipate continued growth in this area of research from both directions, particularly in the following directions.

\subsection{Model Confidence}
The space of possible amino acid combinations is vast, but the space of functional proteins is much smaller, with estimates ranging from one in $10^{11}$ \cite{keefe2001functional} to as low as one in $10^{77}$ \cite{axe2004estimating}. 
This manifold of functional proteins may not have been completely explored by nature, which has search strategies that are biased toward existing functional proteins, and we are still identifying new proteins through metagenomic analysis \cite{mitchell2020mgnify}.
Therefore, it is unclear whether relying on the evolutionary history of identified proteins is sufficient to identify the manifold of functional proteins. 

For protein engineering, it is clear that there is a balance to maintain between exploring new protein sequences and staying within predictive models' confident regions. 
In the directed evolution context, where rounds of experimental data are used to train predictive oracles (or used directly), various methods encode these trust regions heuristically or probabilistically~\cite{brookes2019conditioning,fannjiang2022conformal,biswas2021low,belanger2019biological}, and retraining oracles for specific regions has also been successful~\cite{fannjiang2020autofocused}. 
However, there are few fully enumerated protein landscapes to validate these approaches against.
A nearly comprehensive GB1 study~\cite{wu2016adaptation} has been remarkably useful, as have synthetic datasets, but the field would benefit from more, diverse, epistatic landscapes with which to validate.

While model confidence represents one challenge in protein engineering, noisy labels present another. 
The sample preparation methods used often introduce noise, may have limited linear range of measurement, and can be difficult to reproduce, particularly for newly developed assays designed for to probe more complex biological phenomena. 
While robust training under noisy labels has gained some traction in deep learning, especially for more subjective and error-prone tasks, \cite{natarajan2013learning, song2020learning}, handling noisy labels is typically treated as a data pre-processing step (such as by labeling the data with Gaussian Mixture Models \cite{bryant2021deep, shin2021protein}), although a recent method demonstrates that biophysical priors on functional labels can improve prediction accuracy \cite{nisonoff2022augmenting}.

\subsection{Protein-specific data modalities}
Much of this chapter is focused on the guided acquisition of data for a particular protein function, which is being collected and curated by ProtaBank \cite{wang2018protabank}. 
However, the amount of information available for proteins in other data modalities is also rapidly increasing. 
Many databases exist for various properties of proteins, ranging from protein stability to binding. Of particular note are UniProt \cite{uniprot2021uniprot} and the Protein Data Bank (PDB) \cite{berman2000protein, burley2021rcsb}. 

UniProt contains a large amount of information about proteins beyond their sequence, including cross-references to functional labels, disease-association and Protein family (Pfam) classifications at the per-residue level \cite{mistry2021pfam}, Gene Ontologies on the per-protein example \cite{ashburner2000gene, gene2021gene}, and links to several other databases. 
UniProt also releases reference protein clusters (UniRef \cite{uniprot2021uniprot, steinegger2017mmseqs2}), which are currently almost ubiquitous as the training set for large protein language models \cite{rives2021biological}. 
This dataset will continue to grow as more metagenomes are sequenced \cite{mitchell2020mgnify} and more unique proteins are identified.
Some approaches to modeling proteins are conditioned on protein functional labels \cite{gligorijevic2021function} or other data available in UniProt \cite{madani2020progen}, but there is no clear optimal approach to incorporating annotations about all proteins for protein engineering campaigns, which are often focused on specific protein families.

The PDB is the primary source of protein structure data, and it largely consists of static protein structures obtained through protein crystallography, although the number of structures obtained through cryo-EM and NMR are also increasing. 
Protein structures are invaluable to biologists in providing much-needed context for molecules that are otherwise difficult to probe. 
However, mutations may have effects that are not captured by static structures. 
For example, they may bias the protein's Boltzmann distribution toward different conformational states in the ensemble without perturbing the ground state crystal structure, or the crystal structure may simply have too low resolution to capture small changes. 
Nonetheless, structure can be a useful prior in directed evolution \cite{bepler2019learning, wittmann2021informed}.

While protein data continues to grow, methods with inductive biases tailored for biology are comparatively underdeveloped, particular for protein engineering applications. 
A few notable exceptions have been developed in application to protein structure modeling.
For example, the MSA Transformer \cite{rao2021msa} develops a variant of axial attention specific for MSAs, which contain an aligned homolog sequence in each row. 
The protein sequences in an MSA likely share similar structure, and the MSA transformer is able to leverage this structure by tying row attention maps. 
Famously, AlphaFold \cite{jumper2021highly} incorporated several biophysical inductive biases in the CASP14 protein structure prediction contest. 
These include a variant of attention to account for the triangle inequality on distances, a lowered emphasis on the linear input sequence of a protein (which folds into a three-dimensional structure), and a variant of axial attention for the MSA. 
These methods have been successful for improving protein structure prediction, and it is likely that machine learning methods developed to balance evolutionary history, structure, and function will be successful in protein engineering as well.

\subsection{Summary}
Protein engineering is an optimization strategy, and many recent advances in machine learning have been successfully applied or developed to the directed evolution of proteins. 
In this chapter, we have highlighted studies where machine learning contributes to each step of the optimization process. 
As we continue to gather more protein information and develop methods to bootstrap and guide engineering from these data, machine learning will play an increasingly important role in protein engineering.

\printbibliography

\section{Funding}
This work was funded by
\begin{itemize}
  \item NSF Division of Chemical, Bioengineering, Environmental, and Transport Systems (CBET 1937902)
  \item Amazon AI4Science
  \item Caltech Biotechnology Leadership Program (NIH 5 T32 GM 112592-5)
\end{itemize}

\end{document}